\documentclass[aps,pre,twocolumn,groupedaddress,showpacs,superscriptaddress]{revtex4}

\usepackage{graphicx}
\usepackage{graphics}

\begin{document}
\title{Towards absolute calibration of optical tweezers}
\author{N. B. Viana} 
\affiliation{LPO-COPEA, Instituto de Ci\^encias Biom\'edicas, Universidade Federal do Rio de Janeiro, Rio de Janeiro, RJ, 21941-590, Brazil}
\affiliation{Instituto de F\'{\i}sica, Universidade Federal do Rio de Janeiro, Caixa Postal 68528,   Rio de Janeiro,  RJ, 21941-972, Brazil}
\author{M. S. Rocha}
\author{O. N. Mesquita}
\affiliation{Departamento de F\'{\i}sica, ICEx, Universidade Federal de Minas Gerais, Caixa Postal 702, Belo Horizonte, MG, 31270-901, Brazil}
\author{A. Mazolli}
\affiliation{Instituto de F\'{\i}sica, Universidade Federal do Rio de Janeiro, Caixa Postal 68528,   Rio de Janeiro,  RJ, 21941-972, Brazil}
\author{P. A. Maia Neto}
\affiliation{LPO-COPEA, Instituto de Ci\^encias Biom\'edicas, Universidade Federal do Rio de Janeiro, Rio de Janeiro, RJ, 21941-590, Brazil}
\affiliation{Instituto de F\'{\i}sica, Universidade Federal do Rio de Janeiro, Caixa Postal 68528,   Rio de Janeiro,  RJ, 21941-972, Brazil}
\author{H. M. Nussenzveig}
\affiliation{LPO-COPEA, Instituto de Ci\^encias Biom\'edicas, Universidade Federal do Rio de Janeiro, Rio de Janeiro, RJ, 21941-590, Brazil}
\affiliation{Instituto de F\'{\i}sica, Universidade Federal do Rio de Janeiro, Caixa Postal 68528,   Rio de Janeiro,  RJ, 21941-972, Brazil}

\date{\today}

\begin{abstract} 
Aiming at absolute force calibration of optical tweezers, following a critical review of proposed theoretical models, we present and test the results of MDSA (Mie-Debye-Spherical Aberration) theory, an extension of a previous (MD) model, taking account of spherical aberration at the glass/water interface. This first-principles theory is formulated entirely in terms of experimentally accessible parameters (none adjustable). Careful experimental tests of the MDSA theory, undertaken at two laboratories, with very different setups, are described. A detailed description is given of the procedures employed to measure laser beam waist, local beam power at the transparent microspheres trapped by the tweezers, microsphere radius and the trap transverse stiffness, as a function of radius and height in the (inverted microscope) sample chamber.  We find generally very good agreement  with MDSA theory predictions, for a wide size range, from the Rayleigh domain to large radii, including the values most often employed in practice, and at different chamber heights, both with objective overfilling and underfilling.  The results asymptotically approach geometrical optics in the mean over size intervals, as they should, and this already happens for size parameters not much larger than unity. MDSA predictions for the trapping threshold, position of stiffness peak, stiffness variation with height, multiple equilibrium  points and `hopping' effects among them are verified. Remaining discrepancies are ascribed to focus degradation, possibly arising from objective aberrations in the infrared, not yet included in MDSA theory. 
\end{abstract}
\maketitle

\section{Introduction}

Optical tweezers have become invaluable tools for measuring forces and displacements at the single molecule level in cell biology~\cite{Grier2003}. 
Transparent microspheres are employed as handles and force transducers. In the usual domain of application, the trapping force is well described by Hooke's law, so that force calibration amounts to determining the trap stiffness. Commonly applied methods rely on comparisons with fluid drag forces or on the detection of thermal fluctuation features \cite{NeumanBlock2004}.

Absolute calibration should be based on a theoretical expression for the stiffness in terms of measurable parameters. In the most widely employed optical tweezer setup, an incident TEM$_{00}$ laser beam of vacuum wavelength $\lambda_0$, typically around $1 \mu{\rm m}$,
 is focused by a high numerical aperture oil immersion objective of an inverted microscope through a glass slide onto a transparent microsphere immersed in water inside a sample chamber. One finds that high trapping efficiency requires a beam waist that slightly overfills the objective. The radius $a$ of 
the microsphere usually ranges between $0.1\lambda_0$ and $10 \lambda_0$. 
Thus, an accurate theoretical model should ideally satisfy several requirements:

(i) It should realistically model the strongly focused laser beam produced by the objective. Since most microscope objectives are corrected for the visible, there may be objective aberrations to account for in the infrared.

(ii) With oil immersion, the index discontinuity between the glass slide and the water gives rise to spherical aberration, which degrades the focus and must be included in the beam description.

(iii) The direct interaction between the focused beam and the microsphere spans the full range between Rayleigh scattering and the ray optics limit, so that it should be described in terms of Mie scattering.

(iv) At the upper end of the range of microsphere radii, the results should approach the ray optics approximation. However, as is typical of ``semiclassical" approximations \cite{Nussenzveig1992}, this is not to be understood as a pointwise limit, but rather as a size average over rapid interference oscillations.

(v) The directly scattered beam undergoes multiple reflections between the microsphere and the walls of the sample chamber (reverberation). Usually, only the glass slide may be close enough to have a possibly significant effect.

A brief critical review of proposed theoretical models, as far as possible in chronological order, follows.

A ray-optics model was formulated by Ashkin \cite{Ashkin1992} from preliminary 
results derived by Roosen \cite{Roosen}. It did not take into account that high numerical aperture objectives are designed to satisfy the Abbe sine condition, which was later incorporated into the model by Gu {\it et al.} \cite{Gu1997}.

A widely employed wave-theoretical model is based on so-called GLMT (Generalized Lorentz-Mie Theory), developed by 
Gouesbet {\it et al.}~\cite{Gouesbet1985}.
 The laser beam is represented by an attempted improvement on the paraxial Gaussian TEM$_{00}$ model, 
including fifth-order corrections in powers of the ratio between wavelength and beam waist \cite{Barton1989}.
However, it has been shown~\cite{Ganic2004} that 
such an approximation does not correctly represent the field near the focus of a high NA objective.

A new approach was taken by Tlusty {\it et al.}~\cite{Tlusty1998}. They proposed to approximate the trapping force by the gradient of the dipole interaction energy integrated over the trapped particle, taken with respect to an unperturbed incident Gaussian beam around the focal point. This leads to a Gaussian fall-off of the transverse stiffness as a function of  $a$ 
 for $a \gg \lambda_0$, a result that is incompatible with the hyperbolic fall-off implied by 
dimensional arguments~\cite{MN-Nussenzveig2000}, and which is reflected in the large inconsistency
 they find with experimental data in this region.  

Already in Ashkin's work \cite{Ashkin1992} 
it was remarked that the proper wave description of a highly convergent beam is not a Gaussian, but rather the electromagnetic generalization \cite{RichardsWolf1959} 
of Debye's classic exact scalar representation formulated by Richards and Wolf. 
This representation, with proper accounting for the sine condition, was adopted by Maia Neto and Nussenzveig \cite{MN-Nussenzveig2000}
to evaluate the axial trapping force, with scattering described by Mie theory. 
The force at the geometrical focus shows rapid near-sinusoidal oscillations, accounted for by a simple interferometer picture. The axial stiffness approaches the Rayleigh limit for $k_1 a\ll 1$ ($k_1=2\pi n_1/\lambda_0$ is the wavenumber in the sample region, of index $n_1$)
and shows attenuated interference oscillations for $k_1 a \gg 1$, 
with size average approaching the ray optics result. 

In a new  treatment by Rohrbach and Stelzer~\cite{RohrbachStelzer2001}, an angular spectrum (Fourier) representation of the incident beam \cite{McCutchen1964} analogous to the Richards-Wolf representation is employed. 
However, instead of the exact Mie theory, they extend the approach of Tlusty {\it et al.}
by splitting the trapping force into a gradient force and a scattering force. 
The gradient force is given by an expression similar to that of Tlusty {\it et al.}
 (apart from the modified incident beam). 
The scattering force is obtained in the Rayleigh-Gans approximation, 
in terms of extinction and scattering efficiencies. While the result would suffer from a similar incompatibility with the ray optics limit, this problem does not arise, because use of the Rayleigh-Gans approximation already restricts the domain of applicability to the range 
\begin{equation}\label{RG}
2(n_2-n_1) a\stackrel{<}{\scriptscriptstyle\sim} \lambda_0
\end{equation}
 where $n_2$ is the refractive index of the sphere. 

The MD theory was extended to transverse trapping forces by Mazolli {\it et al.} \cite{Mazolli2003}.
 It was explicitly demonstrated that the exact partial-wave result approaches the ray optics one for large $k_1a,$
 in the sense of a size average, as it should. 
Besides the transverse trap stiffness, the equilibrium position with an external applied force and 
the maximum transverse force were also evaluated, taking due account of the interplay
 between axial and transverse equilibrium. As remarked by Merenda {\it et al.} 
\cite{Merenda2006}, it is the only
wave-theoretical  treatment where this effect was correctly incorporated in the evaluation. 
Only a few experimental points were available for comparisons, and discrepancies were attributed 
to the effects of spherical aberration at the water/glass slide interface [effect (ii) above], not included in the MD theory.

A recent contribution by Rohrbach~\cite{Rohrbach2006}
 does not properly refer to optical tweezers, but to a proposed new instrument, the photonic force microscope 
\cite{Rohrbach2004},
which employs an upright microscope with a water-immersion objective, thereby avoiding this spherical aberration effect. 
The theory is the same as in Ref.~\cite{RohrbachStelzer2001},
 and therefore it is also inconsistent with the ray-optic limit, although the author asserts that it delivers reasonable results even when condition (\ref{RG}) is not satisfied.

An improved version of MD theory, including the effects of spherical aberration at the water/glass slide interface, denoted as MDSA theory and described in Sec. II, was experimentally tested and the results were briefly reported 
by Viana {\it et al.}~\cite{Viana2006a}.  The present work is a detailed presentation and discussion of these results.

\section{MDSA theory}

The effect on a focused beam of spherical aberration produced by refraction at the interface between two 
transparent media on a focused beam, 
extending the Richards-Wolf  solution, has been treated by 
T\"or\"ok {\it et al.} \cite{Torok1995}. 
We follow the same procedure to incorporate this effect into the MD theory.
Since the  refractive index $n$ of the glass slide is larger than the refractive  index $n_1$ 
of the water in the sample chamber,  part of the incident beam angular spectrum of plane waves may exceed  the critical angle. 
 However, we shall neglect  possible contributions from evanescent waves, expected to be  negligible  at distances from the interface larger than the wavelength, as will be assumed. 
Thus, for each plane wave component of the incident beam, the transmission amplitude is

\begin{equation}
T(\theta) = \frac{2\cos\theta}{\cos\theta + N\cos\theta_1},
\end{equation}
where $N=n_1/n,$ $\theta$ is the angle between the wavevector component ${\bf k}$ at the glass slide
(with $k=2\pi n/\lambda_0$) and the 
$z$-axis, and $\theta_1=\arcsin(\sin\theta/N)$ is the corresponding angle in the 
sample chamber~\cite{foot1}.

\begin{figure}[h]
\centering
\includegraphics[width=5cm]{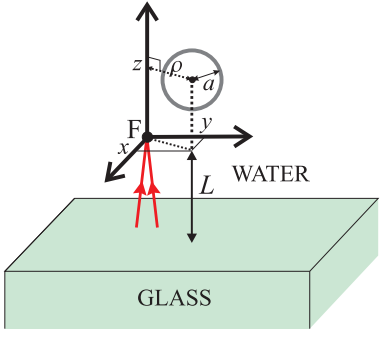}
\caption{Schematic representation of the dielectric microsphere (radius $a$) in the water solution.
The paraxial focus F corresponds to the intersection of paraxial rays. Non-paraxial rays, not represented for clarity, 
intersect the symmetry $z$-axis between F and the glass-water plane interface, which is at a distance $L$ from the paraxial
focal plane. The microsphere position is measured with respect to F. 
}
\label{desenho}
\end{figure}

The refraction also (and more importantly) modifies the phases of the different plane-wave components 
of the laser beam. This is quantified by
the spherical aberration function~\cite{Torok1995} 
\begin{equation}
\Psi(z,\theta) = k\left[-\frac{L}{N}\cos\theta + N(L+z)\cos\theta_1\right],\label{Psi}
\end{equation}
where $L$ is the 
distance between the interface and the paraxial focal plane (Fig.~\ref{desenho}).

The position of the center of the sphere with respect to 
the paraxial focus is $(\rho,\phi,z)$ in cylindrical coordinates. 
For simplicity, we take circular polarization. In this case,  the optical potential 
does not depend on $\phi.$ 
With respect to the ideal case considered 
in Ref.~\cite{Mazolli2003},
 the multipole coefficients of the incident beam are 
modified by multiplying 
each plane-wave component 
by $T(\theta)e^{i\Psi(z,\theta)}:$ 
\begin{equation}\label{G}
G_{jm} =  \int_0^{\theta_0} d\theta \, \sin\theta 
\sqrt{\cos\theta}\, \exp(-\gamma^2\sin^2\theta)\, T(\theta)\,
\end{equation}
\[
\times d^j_{m,1}(\theta_1)\,
J_{m-1}(k\rho\sin\theta)\, e^{i\Psi(z,\theta)},
\]
where $d^j_{m,m'}(\theta)$ are the matrix elements of finite rotations \cite{Edmonds1957}, and 
$J_m$ are the Bessel functions of integer order. The parameter $\gamma = f/w$ is the ratio of the focal length 
to the beam waist at the entrance aperture of the objective ({\it not} at the sample). 
It determines the 
fraction of available beam power that fills the objective aperture (filling factor~\cite{Ashkin1992}). 
The fraction of available power transmitted into the sample chamber is given by
\begin{equation}
A = 16\gamma^2   \label{A}
\int_0^{\sin\theta_0} ds\, s \,\exp(-2\gamma^2s^2)\, \frac{\sqrt{\left( 1 - s^2 \right)\left(N^2- 
s^2\right)}}{\left(\sqrt{1- s^2} + \sqrt{ N^2- s^2}\right)^2}.
\end{equation} 
 In the absence of the interface ($N=1$), $A$
coincides with the filling factor $1-\exp(-2\gamma^2 \sin\theta_0)$ defined in Refs.~\cite{MN-Nussenzveig2000} and \cite{Mazolli2003}. 

Additional transmission losses through the objective, if any, 
must be taken into account to evaluate the actual power $P$ at the sample 
(rather than the power at the entrance aperture)
Investigation of the microscope objective employed 
in the present work has revealed  that its transmittance at the infrared laser wavelength  
is not radially uniform \cite{Viana2006b}. 
This affects not only the power, 
but also the intensity gradients in the sample region. 
 The objective transmission amplitude can be accurately modeled by a Gaussian function:

\begin{equation}\label{Tobj}
T_{\rm obj}(\theta)= T_A \exp\left(-\frac{f^2\sin^2\theta}{4\xi^2}\right),
\end{equation}
where $T_A$ is the axial transmittance amplitude, and $\xi$ 
is a length scale characterizing the objective for a given wavelength.
This effect can readily be incorporated in terms of an effective reduced beam waist 
$w_{\rm eff}$ given by 
\begin{equation}\label{weff}
\frac{1}{w_{\rm eff}^2} = \frac{1}{w^2} + \frac{1}{4\xi^2}.
\end{equation}

We represent the optical force by an efficiency factor in the usual way  \cite{Ashkin1992}:
\begin{equation} \label{QFPc}
{\bf Q}=\frac{\bf F}{n_1 P/c},
\end{equation}
where $P$ is the local laser power at the sample and $c$ is the velocity of light.
The evaluation of the trapping force in terms of a partial-wave series is similar to that described in 
\cite{Mazolli2003}.
We derive two separate contributions: 
${\bf Q} = {\bf Q}_s + {\bf Q}_e.$ 
${\bf Q}_e$ represents the rate of removal of momentum from the incident beam. 
Its axial component is 
\begin{equation}
Q_{{\rm e} z} = \frac{4\gamma^2}{A\,N} \,{\rm Re}\sum_{j,m} (2j+1)(a_j+b_j)G_{j,m}G_{j,m}'{}^{*},
\end{equation} 
where $j$ ranges from one to infinity and $m$ ranges from $-j$ to $j$, 
the star denotes complex conjugation,
$a_j$ and $b_j$ are the Mie coefficients 
\cite{BohrenHuffman1983}, $G_{j,m}$ 
is given by Eq.~(\ref{G}) 
and 
\begin{widetext}
\begin{equation}
G_{j,m}' =  \int_0^{\theta_0} d\theta \, \sin\theta 
\sqrt{\cos\theta}\cos\theta_1\, \exp(-\gamma^2\sin^2\theta)\, T(\theta)\, d^j_{m,1}(\theta_1)\,
J_{m-1}(k\rho\sin\theta)\, e^{i\Psi(z)}.
\end{equation}

The transverse components in cylindrical coordinates are
\begin{equation}
Q_{{\rm e} \rho} = \frac{2\gamma^2}{A\,N} \,{\rm Im}\sum_{j,m} (2j+1)(a_j+b_j)G_{j,m}(G_{j,m+1}^{(-)}
-G_{j,m-1}^{(+)})^*,\label{Qerho}
\end{equation} 
and 
\begin{equation}
Q_{{\rm e} \varphi} = -\frac{2\gamma^2}{A\,N} \,{\rm Re}
\sum_{j,m} (2j+1)(a_j+b_j)G_{j,m}(G_{j,m+1}^{(-)}
+G_{j,m-1}^{(+)})^*,
\end{equation} 
with 
\begin{equation}
G_{j,m}^{\pm} =  \int_0^{\theta_0} d\theta \, \sin\theta 
\sqrt{\cos\theta}\sin\theta_1\, \exp(-\gamma^2\sin^2\theta)\, T(\theta)\, d^j_{m\pm 1,1}(\theta_1)\,
J_{m-1}(k\rho\sin\theta)\, e^{i\Psi(z)}.
\end{equation}
\end{widetext}

${\bf Q}_s$
 represents minus the rate of momentum transfer to the scattered field. Its components are given in the Appendix. 

In this paper, we are particularly interested in the axial component (in order to derive the 
trapping threshold and the stable equilibrium positions, if any), and in the transverse trapping stiffness 
\[
\kappa_{\perp} = -\frac{n_1 P}{c}\frac{\partial Q_{\rho}}{\partial \rho}
\]
where the derivative is taken at the equilibrium position $z_{\rm eq}$, 
which lies along the $z$-axis ($\rho = 0$). 
To obtain $\kappa_{\perp}$, we first perform termwise differentiation of the partial-wave series for 
$Q_{\rho}$  and then perform the summation numerically.
 Similarly, $z_{\rm eq}$ is computed numerically as the root of  
      $Q_z(z_{\rm eq})=0.$ 
When multiple equilibrium positions are found, we take the most stable root, corresponding to the deepest potential well. 

The distance $L$ between the interface and the paraxial focus in Eq.~(\ref{Psi})
 is not experimentally accessible. 
In the experiments, the (inverted) microscope objective is 
first moved down until the trapped microsphere just touches the interface, 
and then the desired height in the sample chamber is obtained by 
moving up the objective through a known distance $d$ 
(for additional experimental details, see next section).  
In order to mimic the experimental conditions, 
we adopt the following procedure. 
We first compute the critical distance $L_c$
 for which the equilibrium position is such that the sphere touches the interface,
 by numerically solving the equation   
$Q_z(z_{\rm eq}=a-L_c)=0$ for $L_c.$
This determines the paraxial focal plane for the initial configuration. 
By moving up the objective through a distance $d,$
 the paraxial focal plane is displaced by $Nd.$ 
Accordingly, we evaluate the equilibrium position and the transverse stiffness taking 
\[
L = L_c + Nd.
\]

\section{Experimental procedures}

\subsection{Experimental setups}

The most important force calibration data for practical applications are the values of the transverse stiffness per unit local power as functions of microsphere radius and height in the sample chamber. Aiming at absolute calibration, we propose to make a blind comparison between theory and experiment, with no fitting parameter. Hence the relevant trap parameters, such as the power $P$
 at the sample and the beam waist $w$ at the entrance aperture of the objective, were measured  directly, and the resulting values were plugged into the MDSA model. Whenever possible, two different techniques were employed for measuring each parameter, 
and the results were checked against each other for consistency. The experiments were performed independently 
at UFMG lab and COPEA lab, reproducibly at intervals of several months. 

\begin{figure}[h]
\centering
\includegraphics[width=6cm]{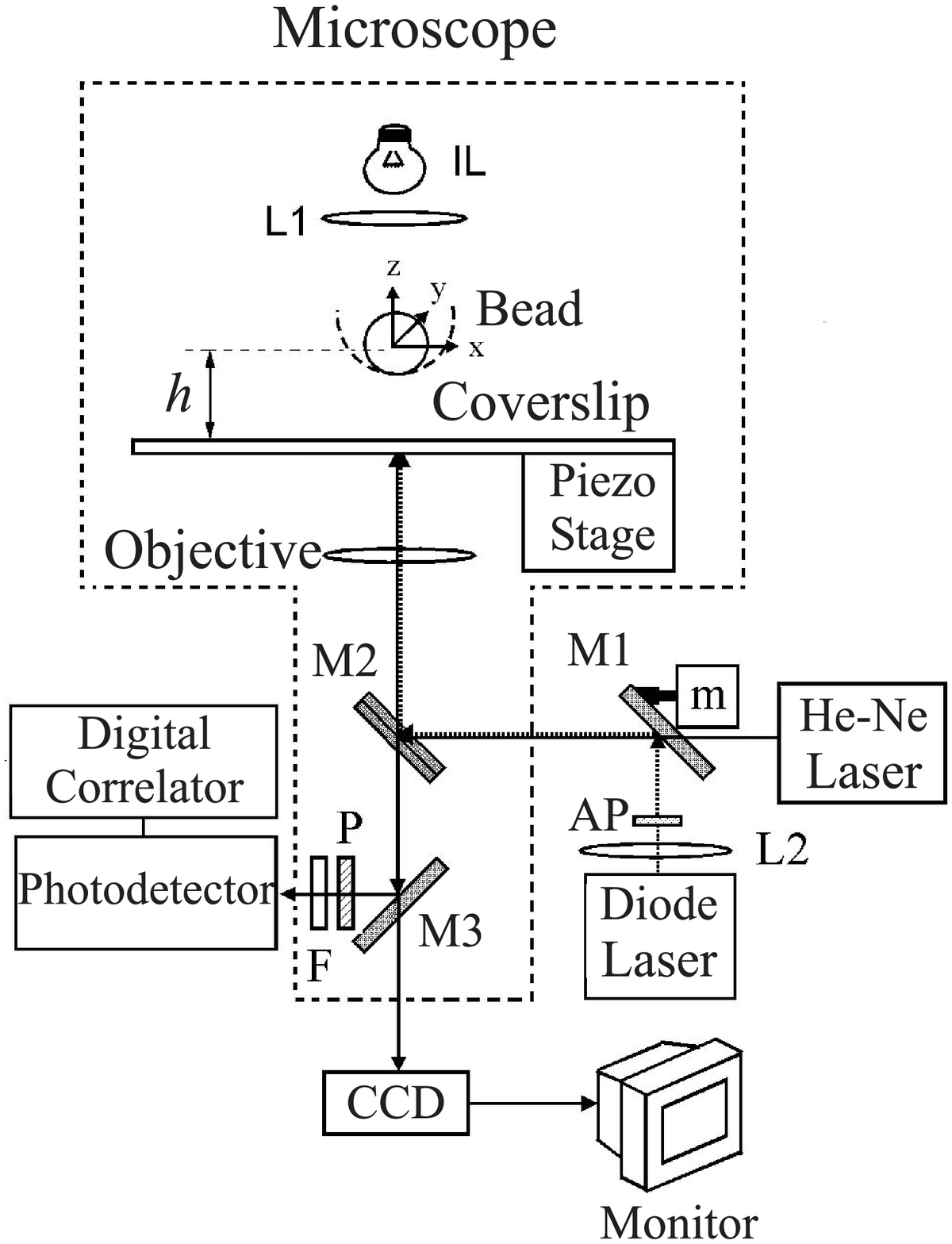}
\caption{Schematic representation of the diode laser setup (DLS).}
\label{setupDLS}
\end{figure}

The two labs employed very different setups, representative of those most often found in practice. Fig.~\ref{setupDLS}
 is a schematic drawing of the Diode Laser Setup (DLS), employed at UFMG lab.

A Nikon TE300 inverted optical microscope (infinity corrected) with 100X N.A. 1.4 Plan Apo CFI objective is employed for trapping, microsphere observation and scattered light collection. To one port is attached  a CCD camera (CCD-72 DAGE-MTI) for visualization; in the other port, we use a photodetector (EGG - Photon Counting Module, SPCM-200-PQ-F500), with collection diameter of 150 $\mu$m, mounted in Newport XY stages for precise positioning. The photodetector delivers TTL pulses ready to be fed into a Brookhaven BI-9000AT digital correlator. An infrared (IR) diode laser (SDL, 5422-H1) operating at 832nm is used for trapping. A He-Ne laser (SP-127), operating at 632.8nm, is the scattering probe. A 20nm width line filter is placed in front of the photodetector to eliminate IR and any light  other than that of the He-Ne laser. A half-wave plate and polarizers are used to control the intensity and polarization of the He-Ne incident and scattered light. A motor (m) is connected to mirror M1, which drives the IR beam onto the objective. The purpose of this motor is to move the IR beam and, consequently, move the trapped bead in relation to the fixed He-Ne laser beam, to obtain the backscattering profile. By determining the backscattering profile and measuring the backscattered light intensity autocorrelation function (ACF), one can obtain the decay time of the Brownian position fluctuations, and finally get the trap stiffness. This procedure is described in detail by Viana {\it et al.}~\cite{Viana2002a}.

\begin{figure}[h]
\centering
\includegraphics[width=7cm]{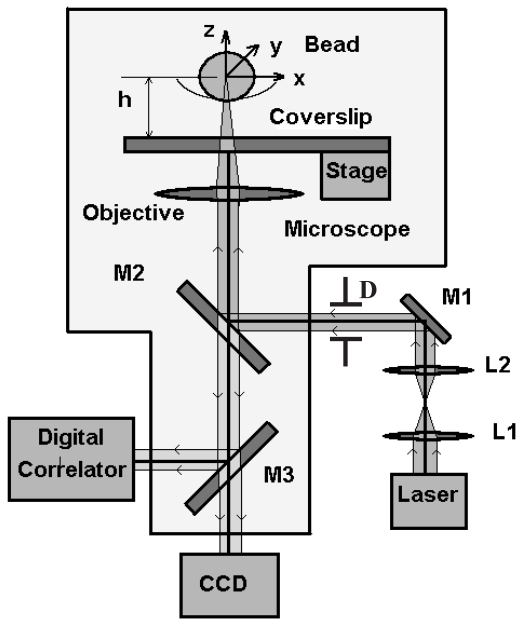}
\caption{Schematic representation of the YAG laser setup (YLS).}
\label{setupYLS}
\end{figure}

The  YAG Laser Setup (YLS), employed at COPEA lab, is shown in 
 Fig.~\ref{setupYLS}.
The YAG laser beam is expanded by the lenses ${\rm L}_1$ and ${\rm L}_2$ to a diameter of $10 {\rm mm}$. 
The microscope and objective are the same model as in the DLS setup. The expanded beam is led to the objective by mirrors  ${\rm M}_1$ and ${\rm M}_2$. Before the microscope epi-illumination port a diaphragm D is employed to control the beam radius. The microscope stage is moved by step motors in the $x$ and $y$ directions, and a Prior motion controller is used to get precise positioning (1 $\mu {\rm m}$ accuracy). We use a CCD camera (Hamamatsu C2400) to observe the samples. In front of the CCD is placed a line filter that blocks light 
at the wavelength 1.064 $\mu{\rm m}$, to prevent scattered and reflected laser light from entering the view field.  The signal from the CCD is fed to a Scion Digital Frame Grabber. The digitized images are analyzed with the NIH ImageJ program.

\subsection{Measurement of beam waist}

To test MDSA theory, it is not necessary to measure the waist at the sample region, which is not an 
input parameter. On the other hand, the beam waist at the entrance aperture of the objective (or more generally
the beam transverse intensity profile) is an important theoretical ingredient, which was measured independently
by two  different techniques: the CCD and the diaphragm methods. 

\begin{figure}[h]
\centering
\includegraphics[width=4cm]{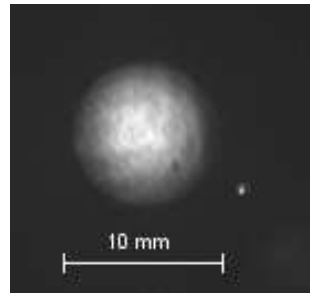}
\caption{YAG laser beam image.}
\label{beamphoto}
\end{figure}

\begin{figure}[h]
\centering
\includegraphics[width=7cm]{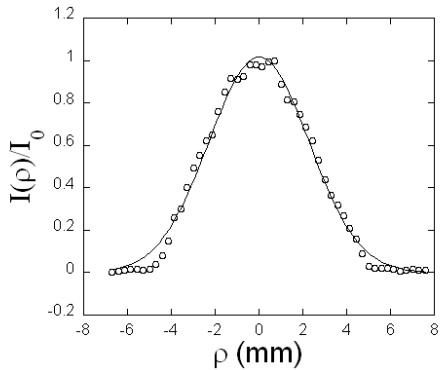}
\caption{
Beam waist measurement  with the CCD method: YAG beam intensity variation in the transverse plane.
The Gaussian fitting (solid line) yields $\sigma_{\rm YLS} = 2.3 \pm 0.2 {\rm mm}$.}
\label{fit}
\end{figure}

In the CCD method, the objective of the microscope is removed and the sample is replaced by a mirror. After taking off the line filter, the reflected beam is imaged by the CCD. To get a ruler calibration,  we first increase the laser power so as to saturate the image and obtain a good definition of the diaphragm border, and then we measure the corresponding diaphragm aperture (see Fig.~\ref{setupYLS}).
Fig.~\ref{beamphoto} shows an image of the expanded YAG beam (10 mm diameter) and Fig.~\ref{fit} shows the radial intensity profile obtained from Fig.~\ref{beamphoto}. The intensity profile is fitted to a Gaussian function,
\begin{equation}
I(\rho)=I_0 e^{-\frac{\rho^2}{2 \sigma^2}}.
\label{n1}
\end{equation}
The beam waist radius is $w = 2\sigma.$ 

From the fit we get $\sigma_{\rm YLS} = 2.3 \pm 0.2 \,{\rm mm}$ for the YAG beam. The error bar was obtained from the statistics of five experiments performed over a period of one year. The same method was applied to the diode laser beam, yielding 
$\sigma_{\rm DLS} = 1.1 \pm 0.1 \,{\rm mm}$.

In the diaphragm method, we vary the diaphragm aperture $R$ 
and measure, as a function of $R$, the transmitted laser power $P(R)$. 
With the Gaussian beam profile fit, we have
\begin{equation}
P(R)=P_t\left(1-e^{-\frac{R^2}{2 \sigma^2}}\right),
\label{n2}
\end{equation}
where $P_t$ is the total power of the beam incident on the diaphragm.

\begin{figure}[h]
\centering
\includegraphics[width=7cm]{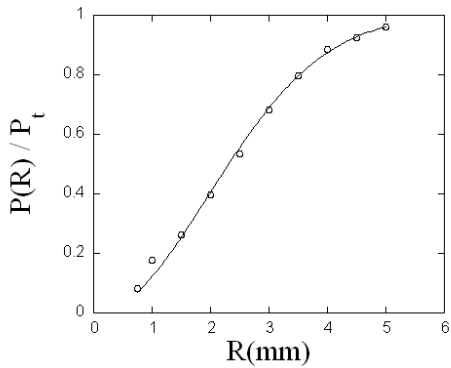}
\caption{
Beam waist measurement  with the diaphragm method:
variation of transmitted power versus   diaphragm radius $R$ for the  
YAG beam.}
\label{PversusR}
\end{figure}

Fig.~\ref{PversusR} 
is a plot of $P(R)$ for the YAG laser. 
Adjusting the measured values to (\ref{n2}), we found a beam half-waist
$\sigma_{YLS}=2.1\pm 0.2 \,{\rm mm}.$
Here again, the error bar was obtained from the statistics of five experiments performed over a period of one year. 
Applying the same method to the diode laser we found a beam half waist 
$\sigma_{DLS}=1.3\pm 0.1 {\rm mm}$.
Therefore, the two methods agree to within the error bars. 

\subsection{Measurement of power at the objective focus}

To measure local power at the sample for the DLS setup, 
two different methods were used: the dual objective method~\cite{Misawa1991}
and a mercury microbolometer method \cite{Viana2002b}.

\begin{figure}[h]
\centering
\includegraphics[width=5cm]{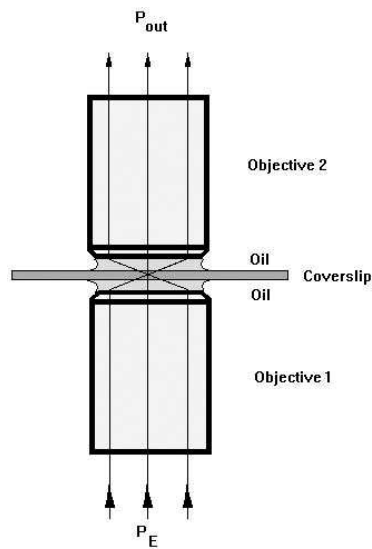}
\caption{Schematic representation of the dual objective method.}
\label{dual}
\end{figure}

Fig.~\ref{dual} 
schematizes the dual objective method. 
On top of the microscope stage, 
with the inverted objective below, is mounted a second, identical objective, in the upright position. 
$P_{\rm E}$ is the power at the entrance of the inverted objective, and $P_{\rm out}$
 is the power transmitted by the compound system. The transmittance is assumed the same for both objectives. The objectives are positioned using three Newport actuators and the microscope stage in order to get a collimated beam emerging
 from the second objective, coaxial with the beam entering the first objective. 

From $P_{\rm E}$ and $P_{\rm out}$ 
we obtain a value for the overall transmittance of the compound system. 
However, care must be taken, since the objective transmittance in
 the infrared is usually not uniform: it is smaller for rays at larger distances from the axis 
(corresponding to larger angles in the sample region) \cite{Viana2006b}. 
Thus, the root-mean-square overall transmittance overestimates the average objective 
transmittance and the power at the sample (see  Ref.~\cite{Viana2006b} for details). 

The Gaussian transmission amplitude model (\ref{Tobj}) 
yields the transmittance function (with $\rho = f \sin\theta$) 
\begin{equation}\label{trobj}
t_{\rm obj}(\rho)= t_A \exp\left(-\frac{\rho^2}{2\xi^2}\right),
\end{equation}
where $t_A=T_A^2$ represents the axial transmittance~\cite{foot_notation}.

Table \ref{t1} shows the
 measured values for the parameters defining the Gaussian transmittance function 
(\ref{trobj}), 
for the YLS and DLS laser wavelengths. 
The entrance aperture radii of the
objectives were $3.5\,{\rm mm}$ for both the DLS and the YLS.
From these values and the laser beam waists at the entrance port,
 we computed the average objective transmittance $t$ for each setup. 
We found $t = 0.20 \pm 0.02$
 for the YLS, and $t = 0.45 \pm 0.03$
  for the DLS. 
As expected, both values are below the respective axial transmittances shown in Table \ref{t1}. 
To evaluate the error bars, we considered the propagation of the errors in the measured beam half waists.

\begin{table}
\caption{\label{t1}Parameters characterizing the Gaussian transmittance function given by
Eq.~(\ref{Tobj}), for two different wavelengths. $t_A$ is the axial transmittance, and $\xi$ is the 
transverse length scale associated to the transmittance radial variation.}
\begin{ruledtabular}
\begin{tabular}{ccc}
 & $t_A$ & $\xi\,({\rm mm})$ \\
\hline
DLS ($\lambda_0 = 0.832\,\mu{\rm m}$) &  $0.53 \pm 0.01$ & $2.2 \pm 0.1$ \\
YLS ($\lambda_0 = 1.064\,\mu{\rm m}$)& $0.31 \pm 0.01$ & $3.3 \pm 0.1$ \\
\end{tabular}
\end{ruledtabular}
\end{table}

\begin{figure}[h]
\centering
\includegraphics[width=5cm]{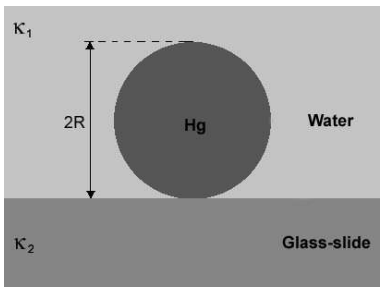}
\caption{Schematic representation of the microbolometer.}
\label{bolometer}
\end{figure}

The value for the DLS setup was checked with the help of a 
 mercury microbolometer  method \cite{Viana2002b}. 
The microbolometer, inserted into the DLS setup, consists of a standard Corning microscope glass slide 
(thickness 170 $\mu{\rm m}$), with an O-ring of 1cm diameter and 0.5cm height glued onto it, filled with water and containing mercury droplets with sizes in the $\mu{\rm m}$ range. 
Since we use an oil immersion objective, the microbolometer does not have any glass-air interface. Moreover,
 the oil refractive index (1.496) being very close to the glass refractive index (1.51), we can ignore refraction at this interface in the analysis. Since the typical radii of the mercury droplets are in the micrometer range, each droplet may be modelled as a sphere embedded within two semi-infinite media (Fig. \ref{bolometer}). 

The incident laser beam is focused by the objective onto the mercury bead, heating it. 
A steady-state situation is achieved  in a time scale of the order of 1 second.
 The mercury heat conductivity is about 13 times larger than the water conductivity. 
Therefore, temperature at the surface of the mercury droplet remains constant for a given  laser power. 
The laser power is then slowly increased until the mercury bead jumps. 
This occurs when its surface temperature reaches the boiling point of water 
(T$_b$ =
97.2$^o$C at the laboratory), making the determination of this point very easy and accurate. 
Knowing the mercury droplet radius, the heat conductivities of the medium 
($\kappa$$_1$ = 0.6791 {\rm W/mK} for 100$^o$C) and
glass slide ($\kappa$$_2$ = 9.43 {\rm W/mK} for 100$^o$C),
the absorption coefficient of mercury for the laser wavelength used (Abs = 0.272), one can determine the local 
 laser power $P.$ 
The objective transmittance is obtained by measuring the mercury droplet radius as a function of the input 
power producing the `jump' (see \cite{Viana2002b} for details).
  We found $t = 0.40 \pm 0.04,$ in agreement with the dual-objective result  within error bars.

\subsection{Measurement of bead radius}

A thorough test of MDSA theory requires 
covering a broad range of microsphere radii. 
Precision calibrated commercial beads are available only for a restricted size range. 
The DLS setup employs uncalibrated immersion oil droplets in de-ionized water,
 produced by a sonicator, yielding an almost continuous range of sizes, from submicrometers to tens of micrometers. The radius of the droplets employed for testing needs to be directly measured.
  
Up to a few $\mu$m, diffraction precludes reliable size determination by videomicroscopy. 
To measure radii in the range from 0.5 to 2.5 $\mu{\rm m}$,
 we employed an alternative  method, 
based on the properties of the free Brownian motion 
 performed by the beads when the optical trap is turned off.
 We infer the hydrodynamic drag coefficient $\beta$
 from the diffusion coefficient $D$
 by using the  Einstein relation, 
and then obtain the radius from Faxen's extension of Stokes's law 
 \cite{NeumanBlock2004}.

To check the method, we first applied it to a deionized water solution of polystyrene 
spheres of radius $a = 1.52 \pm 0.03 \mu{\rm m}.$ A $10^{-4}\%$ 
 solution is placed within a chamber built from an O-ring with $\sim$ 
 1 cm  diameter and $\sim$ 0.3 cm width, 
glued on a coverslip with wax-candle, and covered by another coverslip to avoid evaporation.

To control the bead height $h=z+L$ (see Fig.~\ref{desenho}), 
we first trap it with the optical tweezers and, employing the microscope knob,
 we move it towards the coverslip. When the bead touches the coverslip we see a change in its image. 
This is our  reference height  ($h = a$) 
and we use it to place the bead at a desired height  by acting again on the microscope knob. 
This procedure allows height determination with an uncertainty of 0.5 $\mu$m.

\begin{figure}[h]
\centering
\includegraphics[width=7cm]{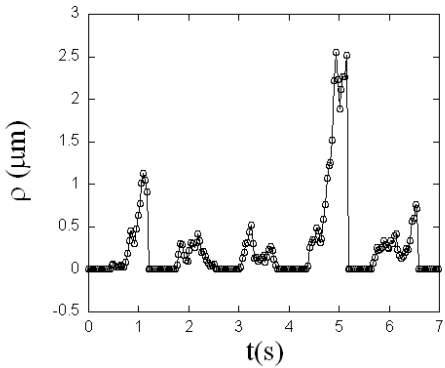}
\caption{Radial distance to the beam axis as a function of time when the optical trap is periodically switched  on and off.
When the trap is off, the bead undergoes free Brownian motion. }
\label{fig8}
\end{figure}

After positioning the bead, we choose a small image area ($\sim$ 50 pixels $\times\sim$ 50 pixels), in order to get a good frame capture rate, 27 to 28 frames per second. 
Then the
position {\boldmath\(\rho\)}
\( =\rho\,\mbox{\boldmath\({\hat \rho}\)}(\phi) \)  of the bead center-of-mass on the xy
plane (the three-dimensional position is \( {\bf r}= \mbox{\boldmath\( \rho \)} + z \mbox{\boldmath\({\hat z}\)}\))
 is
measured at every 1/28 s.
We  turn the optical potential periodically on 
($\sim$ 0.5 s) and off ($\sim$ 0.5 s) by shutting the beam. 
In Fig. \ref{fig8} we show a typical result for the radial distance $\rho(t)$ 
as a function of time. Note that $\rho=0,$ corresponding to the location of the trapping beam, 
also represents the trapping position on the $xy$ plane.

The mean square displacement  is given by
\begin{equation} 
\langle (\delta\rho)^2(t) \rangle = 4 D t,
\label{variance}
\end{equation}
where the diffusion coefficient is given, for temperature $T,$ by the Einstein relation 
($k_B$ = Boltzmann constant)\cite{Furf1956}
\begin{equation} 
D = \frac{k_B T}{\beta}\,.
\label{n12}
\end{equation}
The drag coefficient  $\beta$ (for motion parallel to the glass interface) is given by Faxen's extension of the Stokes law in terms of the bead radius $a$ and the height $h$ \cite{Feitosa1991}:
\begin{equation}  
\beta = \frac{6 \pi \eta a}{ 1-\frac{9}{16}\left(\frac{a}{h}\right)+%
\frac{1}{8} \left( \frac{a}{h}\right)^3-\frac{45}{256}\left(\frac{a}{h}%
\right)^4 - \frac{1}{16}\left(\frac{a}{h}\right)^5}\;.
\label{faxen}
\end{equation}
From the data illustrated in Fig. \ref{fig8}, 
we calculate  $\langle (\delta\rho)^2\rangle,$ excluding the regions where $\delta\rho  < 0.01 \mu{\rm m}$, 
the precision limit for determination of the bead center of mass  using usual centroid-finding algorithms 
\cite{NeumanBlock2004}. Such regions correspond to the time intervals with the trap turned on.  The time interval $t$ in Eq.~(\ref{variance}) is an integer multiple of the inverse  frame acquisition rate $\delta t.$

In view of the time translation symmetry of free Brownian motion, we may combine data from different time intervals, allowing us to improve our statistics in computing $\langle(\delta \rho)^2\rangle.$ 
The data
in Fig. \ref{fig8} represents an ensemble of which each time window with the trap switched off is a realization. For each
realization, we compute the average mean square displacements $\langle (\delta\rho)^2\rangle$  between nearest
neighbors (separated by $\delta t$), between next nearest neighbors (separated by $2\delta t$), and so on,
according to the formula
\begin{equation}
\langle (\delta \rho)^2(k\delta t)\rangle = \frac{k}{N} \sum_{j=0}^{N-k}\left\{\mbox{\boldmath\(\rho\)}[(j+k)\delta t]-
\mbox{\boldmath\(\rho\)}[j\delta t]\right\}^2,
\end{equation}
where $N$ is the total number of frames in a given time window
and  {\boldmath\(\rho\)}
is the radial position vector of the bead center of mass on the $xy$ plane (note that
only \(\rho= | \mbox{\boldmath\( \rho \)} |\) 
is plotted in Fig. \ref{fig8}).

\begin{figure}[h]
\centering
\includegraphics[width=7cm]{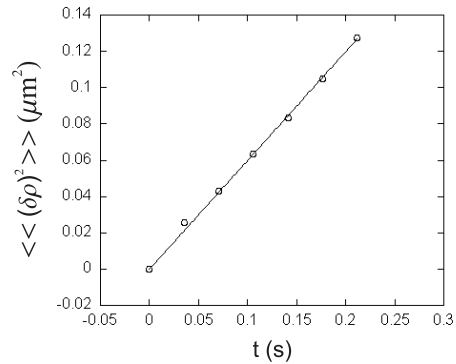}
\caption{Mean square  bead displacement as a function of time.}
\label{fig9}
\end{figure}

The final result $\langle 
\langle (\delta \rho)^2(k\delta t)\rangle \rangle,$ the ensemble average over all time windows, is plotted
in Fig. \ref{fig9} as a function of time $t=k\delta t.$
 From the linear  fitting to (\ref{variance}), we obtain the diffusion coefficient $D$.

\begin{figure}[h]
\centering
\includegraphics[width=7cm]{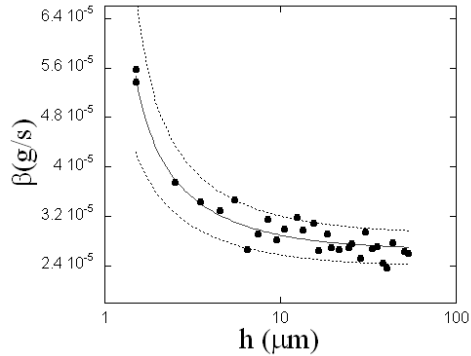}
\caption{Measured drag  coefficient $\beta$ as a function 
of bead height $h.$
 The nominal bead radius is $(1.52 \pm 0.02) \,\mu{\rm m}$, whereas the value measured from the 
Brownian motion is $(1.5 \pm 0.2)\,\mu{\rm m}$}
\label{fig10}
\end{figure}

From the experimental value for $D,$ 
we evaluate the drag coefficient $\beta$ using
Eq.~(\ref{n12}). 
This procedure is repeated at various heights. 
Fig.~\ref{fig10} shows the resulting values of
$\beta$ as a function of $h$. 

The continuous line in Fig. \ref{fig10} 
represents the curve fit to (\ref{faxen}). 
Near the coverslip wall, a small variation in $h$ generates a substantial variation [cf. (\ref{faxen}))]
 in the corresponding value of $\beta$. 
To account for this source of error in height determination, we introduce an off-set $h_0$ 
in the curve fitting function. From the fit we get $h_0=0.3 \mu{\rm m}$ and $a=1.5 \pm 0.2 \mu{\rm m}$. 
The latter is in good agreement with the nominal radius $1.52 \pm 0.02 \mu{\rm m}$ of the calibrated  bead. 

\begin{figure}[h]
\centering
\includegraphics[width=5cm]{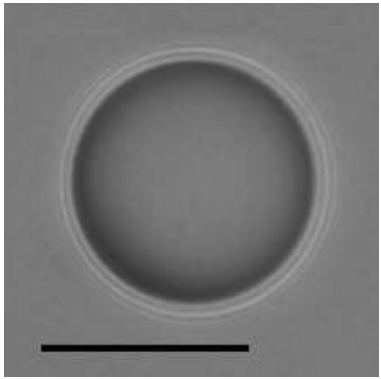}
\caption{Bead image: radius equal to $5.8 \mu{\rm m}$, scale bar 
$10 \mu{\rm m}.$ }
\label{fig11}
\end{figure}

\begin{figure}[h]
\centering
\includegraphics[width=7cm]{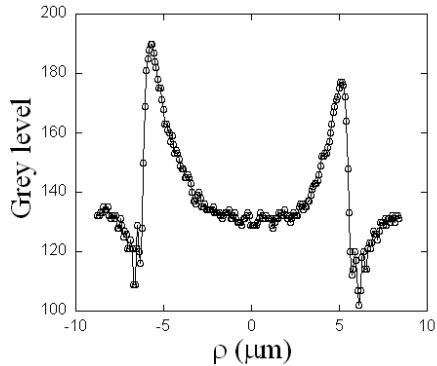}
\caption{Radial grey level of the bead image shown in Fig. \ref{fig11}. 
The bead diameter, defined by the distance between the side hills, equals $11.6\, \mu{\rm m}$.}
\label{fig12}
\end{figure}

The dashed lines in Fig.~\ref{fig10} correspond to the calculated values of $\beta$  for a variation of $\pm 10 \%$ 
in the sphere radius $a.$ 
We see that the data points are spread out between the two dashed lines, showing that the sphere radius is measured with  $ 10 \%$  uncertainty.
For droplets with $a > 2.5 \mu{\rm m},$ the radius was measured by videomicroscopy and ruler calibration. 
Fig.~\ref{fig11} shows the image of a polystyrene sphere of measured radius $5.8\pm 0.1 \mu m$ 
 and Fig.~\ref{fig12} shows the corresponding grey-level profile. We take the sphere diameter to be the distance between the centers of the sigmoid  branches in Fig.~\ref{fig12}, and  the error bar as 1 pixel ($\sim 0.1\, \mu{\rm m}$).

For the YLS setup, we employed a few available 
calibrated beads in the range below $2\,\mu{\rm m},$ and a polydispersion of uncalibrated 
beads above this value, with radii 
measured by videomicroscopy.

\subsection{Measurement of trap stiffness}

\begin{figure}[h]
\centering
\includegraphics[width=7cm]{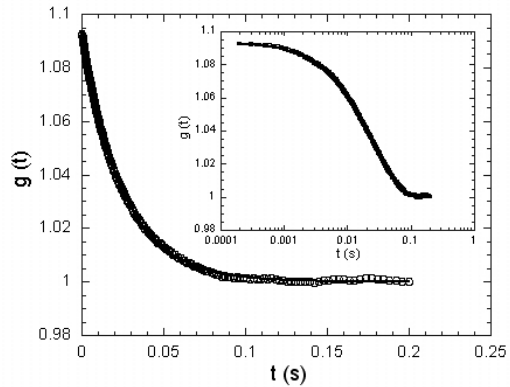}
\caption{Typical normalized ACF fitted to Eq. (\ref{n14}).
\textit{Inset}: same ACF in a semilog plot.}
\label{fig13}
\end{figure}

For the DLS, trap stiffness is measured by analyzing Brownian motion in the optical potential well (as opposed to the free Brownian motion employed for bead size measurement) \cite{Viana2002a}.
A He-Ne laser is used to probe the bead  motion,  by detecting  the intensity fluctuations of backscattered light. By measuring their time intensity autocorrelation function (ACF) with a digital correlator, one obtains the Brownian relaxation time of the bead (see Fig.~\ref{fig13}).

After normalization, the ACF  can be fitted to the following  equation:
\begin{equation}
g(t)= 1 + A_{\perp}\exp(-t/\tau_{\perp}) + A_{z}\exp(-t/\tau_{z})
\label{n14},
\end{equation}
where \textit{A$_{\perp}$} and \textit{A$_{z}$}  are the amplitudes
 and $\tau_{\perp}$ and $\tau_{z}$ are the Brownian motion decay times in the radial and axial directions, respectively. From the resulting value of  $\tau_{\perp}$, we get the transverse stiffness from (see Ref.~\cite{Viana2002a} for details) 
\begin{equation}
\kappa_{\perp} = \frac{\beta}{\tau_{\perp}}\, , \label{n15}
\end{equation}
where the drag coefficient  $\beta$ is known from the measurement of the bead radii (see Sec. III.D). 

The trap stiffness in the YLS was measured using  Faxen's law and videomicroscopy. 
After trapping a bead and positioning it at the desired  height, the microscope stage is set to move laterally with a known velocity $v.$ The Stokes force displaces the bead to a new off-axis equilibrium position. The entire process is recorded, digitizing the images with a Scion frame grabber. The procedure is repeated until one has 5 to 10 different values for the stage velocity in each direction. Velocities are taken small enough to probe only the harmonic sector of the potential well (linear regime). The images are analyzed in order to infer the displacement from the equilibrium position $\delta \rho_{eq}$ as a function of $v,$ checking  that the dependence is linear, of the form
 $\delta \rho_{eq}=\alpha v.$
 The trap stiffness is then obtained from the  angular coefficient $\alpha$ by the relation
\begin{equation}
\kappa_{\perp}=\frac{\beta}{\alpha}.
\label{n16}
\end{equation}

\section{Results}

Following the procedures discussed in Sec. III, we have measured the ratio $\kappa_{\perp}/P$ as a function of 
 $a,$ 
at different heights $h$ above the glass interface. 
Measurements of all relevant data allowed for a direct comparison with MDSA theory, 
with no adjustable parameters. 
In addition to the measurements of beam waist and bead radii discussed in Sec. III, 
we also measured the refractive indices of immersion oil ($n_2 = 1.496$) 
and deionized water ($n_1 = 1.343$) at the diode laser wavelength $\lambda_0=832 \,{\rm nm}$ 
by the minimum deviation method. 

For the polystyrene beads employed in the YLS, we have taken the value  $n_2=1.576$ reported in 
Ref.~\cite{Ma2003}
for  the YAG laser wavelength  $\lambda_0=1064 \,{\rm nm}$, 
and we measured the refractive index of water to be $n_1=1.332$ at this wavelength. 
 Much less important is the refractive index of the  glass slide, which is relevant for the spherical aberration effect, but not for Mie scattering: $n=1.51.$

For the DLS, the objective vertical displacement $d$ 
 was adjusted to have (approximately) the same height $h$ for all bead sizes. 
If the equilibrium position of the bead with respect to the paraxial focus were independent of $L$ (distance between focus and glass wall), we would have a height 
\begin{equation}
h=a+Nd \label{h-d}
\end{equation} 
 after displacing the objective by a distance $d.$ 
A different displacement $d$ was taken for each $a,$ so as to keep a constant $h,$ as given by 
(\ref{h-d}).
Because the optical potential depends on the $L,$ the actual height differs from (\ref{h-d}) by an amount $<a.$ 
This difference is negligible with regard to the Faxen correction of Stokes law (see Section III.D), but it might be important for a quantitative evaluation of the optical force and stiffness. 
Thus, it is taken into account when computing the force and the trap stiffness. 

However, the dependence with  $L$ and $h$ turns out 
to be negligible for the DLS, because the laser beam underfills the objective entrance aperture. 
Thus, the objective produces a more paraxial beam in the sample chamber, rendering the aberration effect irrelevant.  
Indeed, the laser beam waist is rather small ($\sigma_{\rm DLS}= 1.1\pm0.1\,{\rm mm}$, see Sec. III.B) in this setup. 
Moreover, a diaphragm (radius $r_c=2.0\, {\rm mm}$) was employed, cutting off the tails of the transverse intensity profile. This translates into a reduced effective numerical aperture ${\rm NA}_{\rm eff}=
 n \sin\theta_{\rm eff},$
where  $ \sin\theta_{\rm eff} = r_c/f,$ where $f$ is the objective focal length. 
In terms of the entrance aperture  radius  $r_e$, 
the sine condition also yields     $f=r_e/\sin\theta_0,$  
where $\theta_0$ represents the semi-aperture angle for plane-wave illumination.
  Therefore, 
\begin{equation}
{\rm NA}_{\rm eff} = \frac{r_c}{r_e} \,{\rm NA},
\end{equation}
where NA = 1.4 is the objective numerical aperture. 
For our objective, we have $r_e =3.5$ mm, yielding ${\rm NA}_{\rm eff} = 0.8,$ 
which corresponds to $\theta_{\rm eff}=32^o.$ 
Thus, we replace $\theta_0$ by this value in (\ref{G}) and (\ref{A})  when computing the theoretical values for the DLS. 

For the YLS, on the other hand, the aberration effect is important, 
since the laser beam overfills the entrance aperture. 
Plane-wave components at angles $\theta \ge \arcsin(N)$ are refracted into evanescent waves in the sample chamber. 
Since we perform the experiments at a distance of several wavelengths from the glass interface, the contribution of evanescent waves is neglected. When  using the MDSA model presented in Sec.~II, we replace $\theta_0$
 by $\arcsin(N) = 61.9^{\rm o},$ corresponding to an effective numerical aperture ${\rm NA}_{\rm eff}= n_1=1.332.$

As discussed in Sec.~II, 
the radial variation of the objective transmittance is taken into account by using the effective waist defined by 
(\ref{weff}). 
From the values measured for the beam waist (Sec.~III.B) and the length $\xi$ 
characterizing the transmittance variation (Table I of Sec.~III.C), we find $w_{\rm eff}= 2.2$ mm and 
$w_{\rm eff}= 3.5$ mm for the DLS and the YLS, respectively. 

\subsection{Trap stiffness}

\begin{figure}[h]
\centering
\includegraphics[width=7cm]{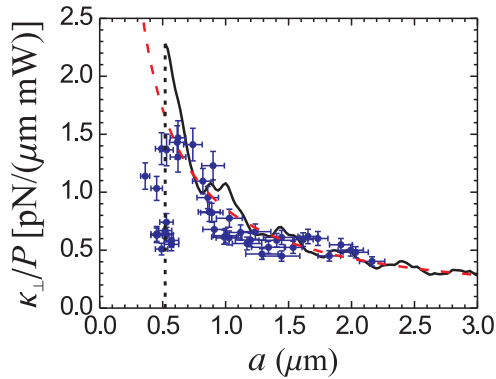}
\caption{Transverse trap stiffness (divided by the local power) as a function of bead radius for oil beads ($n_2 = 1.496$)
 in deionized water ($n_1 = 1.343$) at height $h = 3.1 \,\mu{\rm m},$ for the DLS
 ($\lambda_0 = 0.832\, \mu {\rm m},$ objective underfilling). 
Circles: experimental points (with error bars). 
Solid line: MDSA theory. 
Dashed line: geometrical optics. 
The threshold for trapping is indicated by a dotted vertical line.}
\label{prlfig1}
\end{figure}

Fig.~\ref{prlfig1} displays the results for the DLS  
at \mbox{$h = (3.1 \pm 0.5) \,\mu{\rm m}.$} 
Data points are
averages of 4 independent measurements (each one lasting 100s) for every microsphere;
vertical error bars show the associated standard deviations.
Full and dashed lines correspond to MDSA and GO (geometrical optic, i. e., WKB)  theories, respectively. 
For   $0.04\, \mu{\rm m} < a < 0.52 \, \mu{\rm m},$ 
the former predicts that no stable trapping is possible, because no equilibrium position is found in this 
range~\cite{foot2}. This is indicated by the vertical dotted line break. Correspondingly, experimental points cluster around different values in the neighborhood of the threshold at $a=0.52\,\mu{\rm m}.$
In this range,
the microspheres often escaped from the trap during the measurement interval. Scattered
data points closely below threshold arise from microspheres that stayed in the trap for at
least 3 measurements (around 300 s). Since the harmonic potential well approximation breaks
down in this metastable region, such data should be regarded mainly as confirmation of
unstable below-threshold behavior.

As shown analytically in Ref.~\cite{Mazolli2003}, the GO values correspond to size averages over the oscillations of the exact curve  for sufficiently large values of the size parameter $2\pi n_1 a/\lambda_0.$  
Fig.~\ref{prlfig1} shows that GO still provides not too bad estimates of stiffness for sizes down to $a\sim \lambda_0.$  This is not surprising, since incident rays that contribute when the sphere center is axially located are unaffected by the dominant diffraction and resonance effects in Mie scattering, which are most important for rays incident near the edge of the microsphere, and WKB is a good approximation down to size parameters of order unity outside of this edge domain \cite{Nussenzveig1992}. Prior
estimates of the domain of applicability of GO \cite{NeumanBlock2004} placed it beyond
size parameter 30 ($a > 5\lambda_0$ ). Our present theoretical and experimental extension of that
domain has obvious practical relevance.

\begin{figure}[h]
\centering
\includegraphics[width=8cm]{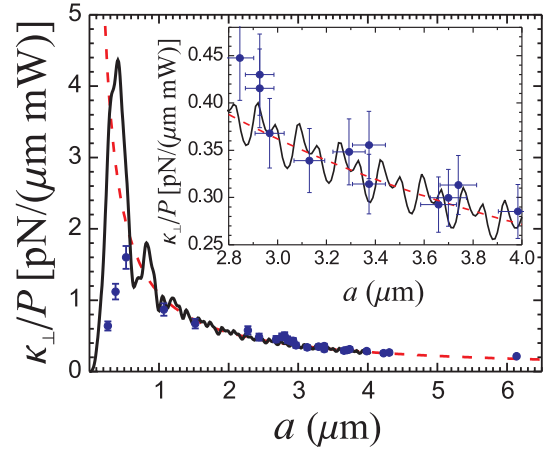}
\caption{Same conventions as in Fig. \ref{prlfig1}, 
for polystyrene beads ($n_2 = 1.576$) 
in deionized water ($n_1 = 1.332$) at  $d = 3\,\mu{\rm m},$ for the YLS 
 ($\lambda_0 = 1.064 \,\mu{\rm m},$ moderate objective overfilling).
  Inset: amplification of the region $2.8 \,\mu{\rm m} \le  a \le  4.0 \,\mu{\rm m}.$}
\label{prlfig2}
\end{figure}

Fig. \ref{prlfig2} shows the results for the YLS setup with $d=(3.0\pm 0.5) \,\mu{\rm m},$ 
 which corresponds to sphere heights  $h - a \approx (2.7\pm 0.5) \,\mu{\rm m}$
according to Eq.~(\ref{h-d}).
 Polydispersions allowed taking many data points for larger $a,$
 but only a few (calibrated) polystyrene sizes were available for smaller $a.$ 
For
such microspheres, at least 4 independent measurements were taken (vertical bars represent the corresponding standard deviations).
Here we find (both experimentally and theoretically) stable trapping below the peak of the stiffness curve. As compared to the DLS, the major difference is the use of (moderate) objective overfilling in this setup.

The amplified scale in the inset of Fig. \ref{prlfig2}  reveals the persistence of oscillations around the GO curve  in the tail of the theoretical curve. They arise from interference among multiple reflections at the upper and lower microsphere interfaces. As shown in \cite{MN-Nussenzveig2000}, 
these oscillations can be derived analytically from the partial-wave series. The corresponding period 
is given by $\Delta a = \frac{1}{4} \lambda_0/n_2\approx 0.169 \,\mu{\rm m},$
  in very good agreement with the solid line curve  in the inset of  Fig. \ref{prlfig2}.
 For the DLS setup, on the other hand,  the oscillations are distorted (Fig. \ref{prlfig1})
 by the beam aperture constraint.  

Our measurement accuracy is not sufficient to verify the presence of the interference
oscillations, a very demanding experimental challenge. However, the inset shows that theory
already predicts a band of halfwidth of the order of $5\%$, about half our measurement
uncertainty, around the GO average, within which data points are expected to fall.

For the lowest measured $a,$ theory predicts two equilibrium points, but only the most stable one, closer to the paraxial focus, is considered. We have checked the numerical results against the Rayleigh limit (very small $a$). In this regime, the force is proportional to the intensity gradient of the incident field alone, allowing us to compare our results with those of ~\cite{Sheppard1997}. 
Spherical aberration leads to the appearance of two local intensity maxima along the axis, which correspond to stable equilibrium positions in the Rayleigh limit. Outside the Rayleigh regime, but with small values of $a,$  there are still multiple equilibria. This is clearly connected with the spherical aberration effect (thus explaining why no such effect takes place in the DLS). 

\begin{figure}[h]
\centering
\includegraphics[width=8cm]{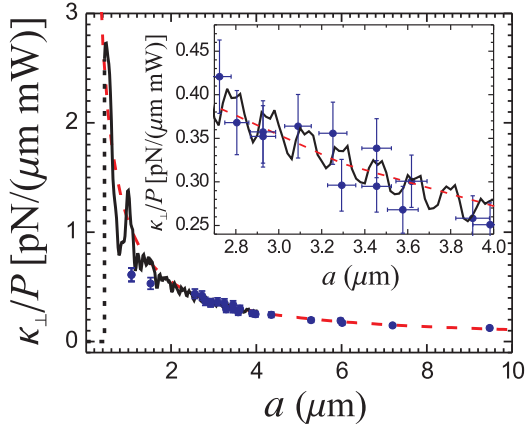}
\caption{Same as Fig. \ref{prlfig2}, at $d = 15 \,\mu{\rm m}.$}
\label{prlfig3}
\end{figure}

\begin{figure}[h]
\centering
\includegraphics[width=7cm]{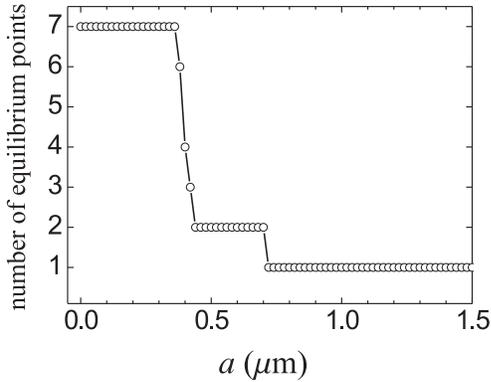}
\caption{Number of stable equilibrium points as a function of sphere radius, for the YLS with $d=15\,\mu{\rm m}.$}
\label{number}
\end{figure}

We may enhance the effect of spherical aberration  by increasing the vertical displacement $d$ of the objective. 
In Fig. \ref{prlfig3}, we show the YLS results for $d=15\,\mu{\rm m},$ 
 corresponding to $h - a \approx (13.2 \pm 0.5) \,\mu{\rm m}.$
 The inset shows that the period of oscillation for large $a$ is not modified by the aberration effect, as expected. In the Rayleigh regime, on the other hand, we find seven stable equilibrium points. They correspond to local intensity maxima lying along the axis, between the glass slide and the paraxial focus, that result from interference fringes bordering the focal region in the diffraction theory of spherical aberration~\cite{Sheppard1997}.
 As we decrease $a,$ starting from the value   $a = 0.44\,\mu{\rm m},$
 the number of equilibrium points rapidly increases and then saturates in the Rayleigh regime, as shown in 
Fig.~\ref{number}. 
 In this regime of multiple equilibria, it is very difficult to achieve trapping and to measure the stiffness. Moreover, the potential well around the equilibrium point near the paraxial focus is usually very shallow in this case (see for instance the solid line in Fig.~\ref{prlfig4} for $a=0.265\,\mu{\rm m}$). Thus, spherical aberration degrades trapping efficiency for small $a,$ producing a `threshold' for stable trapping at $a=0.44\,\mu{\rm m}.$  This is indicated by the vertical dotted line in Fig.~\ref{prlfig3} (further discussion of multiple equilibrium  points is given in Sec.~IV.B). 
By comparing this figure with Fig.~\ref{prlfig2}, we conclude that
spherical aberration decreases the peak stiffness value and slightly displaces the peak towards larger values of $a.$ 
On the other hand, the effect  on stiffness is very small for larger $a$ values   (GO limit).

Increasing the height above the glass slide does not change the stiffness in the DLS, 
as expected for a more paraxial beam. Measured and computed stiffness results at 
$h = (8.6 \pm 0.5) \,\mu{\rm m}$  (not shown) differ very little from those of Fig.~\ref{prlfig1}.

\subsection{Bead hops}

As discussed in Sec. IV.A, the optical potential for small sphere radii contains several equilibrium  points because of the spherical aberration effect. In the course of Brownian motion of the bead, it may hop from some local minimum of the optical potential to a  more stable neighboring minimum.  We have observed this effect and measured the size of the hop. 

\begin{figure}[h]
\centering
\includegraphics[width=7cm]{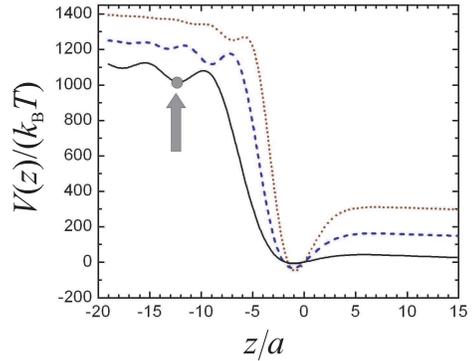}
\caption{Optical potential along the $z$-axis (in units of $k_BT$) 
for the YLS, with $a=0.265 \,\mu{\rm m}$  and  local power $P = 20 {\rm mW}.$
 Objective displacement: $d = 15 \,\mu{\rm m}$
 (solid line),  $ d = 6.5 \,\mu{\rm m}$
  (dashed line) and $d = 3.0 \,\mu{\rm m}$
(dotted line).  The arrow points to the most stable equilibrium point for $d = 15 \,\mu{\rm m}.$
}
\label{prlfig4}
\end{figure}

In Fig.~\ref{prlfig4}, we plot the optical potential along the $z$-axis in units of $k_B T$ 
 as a function of $z/a,$ for three different values of the objective vertical displacement: 
$d = 15 \,\mu{\rm m}$ (solid line),  $d = 6.5 \,\mu{\rm m}$ (dashed line) and $d = 3.0 \,\mu{\rm m}$   
(dotted line). We take parameters corresponding to the YLS,  with  $a=0.265\,\mu{\rm m}$ and $P = 20\, {\rm mW}.$

Besides the equilibrium point nearest to the paraxial focus ($z = 0$), Fig.~\ref{prlfig4} 
shows additional equilibrium points, located between  $z = 0$  and the glass slide. 
They decrease in number and in degree of stability (as measured by the corresponding well depth) as $d$ decreases. 

The arrow in  Fig.~\ref{prlfig4}  
indicates the most stable equilibrium position when $d = 15 \,\mu{\rm m}.$ 
As $d$ decreases  to $6.5 \,\mu{\rm m}$, its  well depth remains close to  $60k_B T$, and then starts to 
decrease much faster below this point (this local minimum also approaches the paraxial focus, as expected, because the interval between this point and the  glass slide, which contains all local intensity maxima,  becomes shorter as $d$ decreases). 
At $d = 6.5 \,\mu{\rm m}$, the potential well near the focus is already much deeper,  and the distance between the two equilibrium points is $2.1 \,\mu{\rm m}.$

\begin{figure}[h]
\centering
\includegraphics[width=7cm]{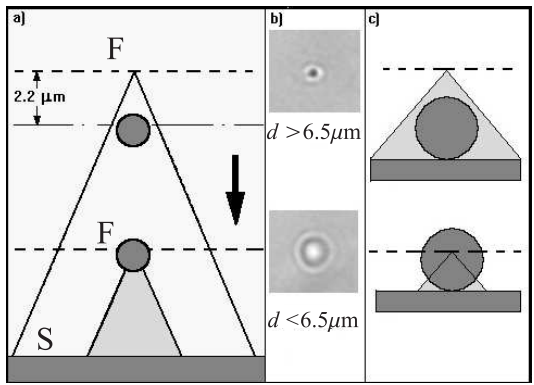}
\caption{
Bead hop in the optical potential.
{\bf a)} Schematic representation of the position of the trapped bead,
paraxial focus F (origin of the $z$ axis) and glass slide S for $d > 6.5 \mu m$ (top, bead below F) and 
$d < 6.5 \mu m$ (bottom, bead close to F). 
{\bf b)} 
Bead images for $d > 6.5 \mu m$ (top) and $d < 6.5 \mu m$ (bottom). 
{\bf c)} Calibration of bead-focus distance with bead attached to the glass slide. }
\label{fig16}
\end{figure}

These predictions were tested by video microscopy, following the image of the trapped bead as the objective was displaced downwards (thus bringing the paraxial focus closer to the glass slide), from $d = 15 \,\mu{\rm m}$ 
down to $d = 3.0 \,\mu{\rm m}.$  The image sharpness depends essentially on the bead position with respect to the paraxial focus. Initially, the bead  is trapped at the position indicated by the arrow in Fig.~\ref{prlfig4}.
  As we bring down the objective, at first the image does not change, indicating that the bead follows the displacement of the paraxial focus, keeping an approximately constant distance to this point (`tweezer' effect). 
Hence, in this  first stage the bead remains in the same potential well.
 However, at $d = 6.5 \,\mu{\rm m}$, we observed a sudden change in the image (see Fig. \ref{fig16}). 
Since the bead size is a fraction of the wavelength, the
image behaves approximately like the diffraction pattern of a point source by the circular objective aperture.
On the focal plane (bottom image), this approaches the standard Airy pattern. Away from the
focal plane (top image), it corresponds to a Fresnel diffraction pattern, with central spot
brightness (in this case, dark) related to the number of Fresnel zones within the aperture
\cite{Jenkins}. 

In order to calibrate the  bead-focus distance, we repeated the experiment with the bead attached to the glass slide. In this case, the distance between bead and paraxial focus is known for each position of the objective. By comparing the resulting images with those obtained with the optically trapped bead, we derive the length of the (upward) hop to be 
$2.2 \pm 0.5 \mu m$, in good agreement with the MDSA prediction for the distance between the two equilibrium points at   $d = 6.5 \,\mu{\rm m},$ shown in Fig.~\ref{prlfig4} (dashed line).

When the objective is moved upwards, we do not observe a reverse hop. Instead, the initially trapped bead  is lost as $d$ approaches $15 \,\mu{\rm m}$. This is also consistent with the solid line in Fig.~\ref{prlfig4}: 
as the well near the focus becomes shallower, the bead cannot climb the optical potential barrier and reach the deeper well below the focus (indicated by the arrow in Fig.~\ref{prlfig4}). This is in line with our previous discussion about the difficulty of trapping small  beads when the paraxial focus is too far from the glass slide (`threshold' of 
Fig.~\ref{prlfig3}).   

\section{Conclusions}

We have demonstrated that the MDSA {\it ab initio} theory of trapping forces in optical tweezers, formulated solely in terms of experimentally accessible parameters, already comes close to achieving a basis for absolute calibration. 
Indeed, we have verified in detail its consequences concerning trap stiffness and trapping thresholds for two very different setups, peak locations, relationship with and domain of validity of geometrical optics, size and height
 dependence resulting from interface spherical aberration, covering the whole range from the Rayleigh regime to the geometrical optic one, and including the description of multiple equilibria and `hop' effects. 

	Among many attempts (Sec.~I), it is the only treatment that satisfies requirement (iv) of Sec.~I, asymptotically approaching geometrical optics in the mean. This is specially important in view of our finding that geometrical optics is already a reasonable approximation, as far as trapping stiffness is concerned,
 for radii of the order 
of the wavelength (a consequence of the absence of near-edge diffraction effects). 

The combined experimental procedures and experimental results reported here, with well-defined and accurately measured parameters, far outnumber all previously available optical tweezers stiffness data. Comparison with MDSA theory shows that, within the typical $10\%$  order of magnitude of the error bars, theory and experiment are generally in very good agreement, for a broad size range and two very different (and often employed) setups, one with underfilling and the other one with moderate overfilling of the objective. The theory accurately predicts the trapping threshold and the location of the stiffness peak, which is extremely sensitive to beam shape, as well as the effects of spherical aberration arising from refraction at the interface,  including multiple equilibria situations. 

The largest deviation from theoretical predictions occurs within the stiffness peak, where the steeply changing curve is most sensitive to additional perturbations, for the YAG laser setup. 
As the microsphere radius decreases in this region, one enters the domain of multiple equilibria and Brownian fluctuations are enhanced. Note (Fig.~\ref{prlfig2}) that the experimental result lies below the MDSA prediction, indicating that missing effects presumably arise from degradation of the focal region. Natural candidates are objective aberrations in the infrared  [effect  (i) in Sec.~I]. In fact, agreement is better for the diode laser setup, 
which employs a more paraxial beam  and a wavelength closer to the 
visible range. 

Other effects not taken into account in MDSA theory include reverberation (multiple beam reflections between the bead and the glass slide) and the effects of evanescent waves beyond the critical angle, as well as possible bead surface distortions or contaminations. In applications requiring accurate force measurements, in view of these perturbations, it seems advisable to stay at least a couple of wavelengths away from the glass slide. Accounting for the deviations may yield new insights concerning bead-surface interactions at close range. 

Polarization effects were not discussed in the present work: 
we found the  transverse stiffness to be  practically independent of polarization.
Polarization effects might be relevant, on the other hand,
when the sphere is at a distance from the axis larger than or of the order of the radius \cite{Fontes2005}.  
They will be discussed in a forthcoming work. 

It has sometimes been argued  that absolute calibration is an impossible aim, in view of the multiplicity of effects that need to be taken into account, as well as the errors in parameter determination. We do not share this pessimistic outlook. Mie scattering, the basic interaction involved, is understood and verified at a level of precision approaching that of quantum electrodynamics \cite{Nussenzveig1992}. 
While we had to employ uncalibrated microspheres to cover a wide size range, NIST-traceable calibrated microspheres are commercially available.  Achieving an understanding and control over additional perturbing effects can turn out to be relevant not only to improving the performance of conventional optical tweezers, but may have potential applications to a variety of new techniques in precision microscopy.

\acknowledgments

This work was supported by the Brazilian agencies Conselho Nacional de Desenvolvimento Cient\'{\i}fico e Tecnol\'ogico
 (CNPq), Funda\c c\~ao  de Amparo \`a Pesquisa de Minas Gerais (FAPEMIG), Projeto Mil\^enio de Nanotecnologia,
 Projeto Mil\^enio de 
Informa\c c\~ao Qu\^antica, Funda\c c\~ao de Amparo \`a Pesquisa do Rio de Janeiro (FAPERJ) and Funda\c c\~ao 
Universit\'aria Jos\'e Bonif\'acio (FUJB). M.S.R. acknowledges support by Laborat\'orio Nacional de Luz S\'{\i}ncrotron (LNLS). 

\appendix
\section{Momentum transfer rate to the scattered field}

In this appendix, we write the explicit partial-wave series for the cylindrical components of ${\bf Q}_s,$
which represents minus the momentum transfer rate to the scattered field:
\begin{widetext}
\begin{equation}
Q_{{\rm s} z} = -\frac{8\gamma^2}{A\,N} \,{\rm Re}\sum_{j,m} 
\left[\frac{\sqrt{j(j+2)(j-m+1)(j+m+1)}}{j+1} \left(a_ja_{j+1}^* + b_j b_{j+1}^*\right) G_{j,m}G_{j+1,m}^*
+ \frac{2j+1}{j(j+1)}ma_jb_j^*|G_{j,m}|^2\right],
\end{equation} 
\begin{eqnarray}
Q_{{\rm s} \rho} = & \frac{4\gamma^2}{A\,N} \,{\rm Im}\sum_{j,m} 
\Biggl[\frac{\sqrt{j(j+2)(j+m+1)(j+m+2)}}{j+1} \left(a_ja_{j+1}^* + b_j b_{j+1}^*\right) \left(G_{j,m}G_{j+1,m+1}^*
+ G_{j,-m}G_{j+1,-m-1}^*\right)\label{Qsrho} \\
 & -2\frac{2j+1}{j(j+1)}\sqrt{(j-m)(j+m+1)} {\rm Re}(a_jb_j^*) G_{j,m}G_{j,m+1}^*\Biggr],\nonumber
\end{eqnarray} 
and
\begin{eqnarray}
Q_{{\rm s} \varphi} = &- \frac{4\gamma^2}{A\,N} \,{\rm Re}\sum_{j,m} 
\Biggl[\frac{\sqrt{j(j+2)(j+m+1)(j+m+2)}}{j+1} \left(a_ja_{j+1}^* + b_j b_{j+1}^*\right) \left(G_{j,m}G_{j+1,m+1}^*
- G_{j,-m}G_{j+1,-m-1}^*\right) \\
 & -2\frac{2j+1}{j(j+1)}\sqrt{(j-m)(j+m+1)} {\rm Re}(a_jb_j^*) G_{j,m}G_{j,m+1}^*\Biggr].\nonumber
\end{eqnarray} 
\end{widetext}


\begin{thebibliography}{999}
\bibitem{foot2} On the other hand, an equilibrium position very close to the focus is found for any radius 
in the range
$a<0.04\,\mu{\rm m},$ as expected for the Rayleigh regime. Of course, this is of little practical value.
\end{thebibliography}
\end{document}